\documentclass[aps,prl,superscriptaddress,amsmath,amssymb,floatfix,twocolumn,showpacs]{revtex4}

\usepackage{amsmath, dcolumn, bm, amsthm, amsfonts, amssymb, tabularx, slashbox, subfigure}
\usepackage{hyperref}
\usepackage{times}
\usepackage{graphicx,graphics,color}

\begin{document}

\title{Coexistence of Enhanced Superconductivity and Antiferromagnetism: Possible Correlated Phase Transitions in Trilayer High-$\textit{T}_\textit{c}$ Cuprates}

\author{Chun Chen}
\email[Corresponding author: ]{cchen@physics.umn.edu}
\affiliation{School of Physics and Astronomy, University of Minnesota, Minneapolis, Minnesota 55455, USA}

\author{A. Fujimori}
\affiliation{Department of Physics, University of Tokyo, Bunkyo-ku, Tokyo 113-0033, Japan}

\author{C. S. Ting}
\affiliation{Texas Center for Superconductivity and Department of Physics, University of Houston, Houston, Texas 77204, USA}

\author{Yan Chen}
\email[]{yanchen99@fudan.edu.cn}
\affiliation{Department of Physics, Surface Physics Laboratory (National Key Laboratory) and Laboratory of Advanced Materials, Fudan University, Shanghai 200433, China}

\date{\today}

\begin{abstract}

Based on a hybrid interlayer coupling mechanism, we study the coexistence of superconductivity (SC) and antiferromagnetism (AFM) in trilayer cuprates. By introducing an interlayer magnetic scattering term, we solve the multilayer $t\mbox{-}J$ model with Josephson coupling under the framework of Gutzwiller projection. We show that both the SC and AFM orders in the multilayered system are enhanced and the range of AFM order is extended. The layer configuration of $d$-wave pairing gap and AFM order further plays an essential role in determining the interlayer magnetic and superconducting coupling phase diagram of such multilayered systems. Abrupt phase transitions between correlated states carrying distinct configurational symmetries are unveiled by tuning the doping level and/or the tunneling strengths.

\end{abstract}

\pacs{74.78.Fk, 74.72.Gh, 74.25.Dw}

\maketitle

{\it Motivation.}---Strong electron correlations and unconventional order parameters fuse together to engender a plethora of fascinating phenomena in multilayered high-$\textit{T}_\textit{c}$ cuprate superconductors (HTSC). Among them is the discovery of the coexistence of HTSC phase with the AFM metallic phase in site-selective nuclear magnetic resonance (NMR) studies of multilayer cuprates \cite{mukuda,shimizu1,shimizu2}. Experimentally, the presence of outer CuO$_{2}$ layers provides the effective screening of the random potential influence from the charge reservoirs in the inner CuO$_{2}$ planes. This intrinsic feature greatly facilitates the realization of the coexistence of HTSC with AFM in multilayered cuprates and makes their mutual interplay a central issue as compared to single layered systems. Furthermore, a striking common aspect shared by the phase diagrams of a large class of SC materials including cuprates, iron-pnictides and several heavy-fermion superconductors is the close proximity of SC phase with the AFM or spin-density-wave phase \cite{scalapino}. Therefore, such an intriguing mixture of HTSC and AFM in multilayered copper oxides sets up a new platform for investigating the relationship between these two orders. It also opens the key avenue toward uncovering the mechanism underlying HTSC in cuprates and related materials.

Angle-resolved photoemission spectroscopy (ARPES) measurements on the high-quality optimally doped triple-layer cuprate superconductor $\textrm{Bi}_{2}\textrm{Sr}_{2}\textrm{Ca}_{2}\textrm{Cu}_{3}\textrm{O}_{10+\delta}$ (Bi$2223$) further reveal a layer variation of both doping density and $d$-wave SC gap through successfully observing the electronic structure and multilayer band splitting \cite{ideta}. In particular, two symmetric outer CuO$_{2}$ planes (OPs) are overdoped with gaps which are larger than those for optimally doped single layer cuprates while the inner CuO$_{2}$ plane (IP) is underdoped with an even larger gap. Generally, an interlayer Cooper pair tunneling has always been put forward as an approach to increase the SC critical temperature of multilayer cuprates \cite{wheatley1,wheatley2} and was extensively elaborated to explain why $\textit{T}_\textit{c}$ takes maximum at $n\!\!=\!\!3$ \cite{chakravarty1}. Nevertheless, there is still a significant question concerning the role of Cooper pairs tunneling in the AFM background when the uniformly coexistent AFM-SC state was discovered in multilayered copper oxides.

Theoretically, although it is known that two-dimensional $t\mbox{-}J$ model allows the emergence of AFM order and its coexistence with HTSC in the underdoped region \cite{ogatarev}, the underlying mechanism of enlarging the coexisting range between AFM and HTSC phases in multilayer cuprates is still under controversial debate \cite{wqchen,mori1,mori2,gan,yamase,yoshizumi}. To this end, a proper understanding of the exhibited broadening of the AFM-SC mixed regime remains much needed. Moreover, both SC and AFM orders carry nontrivial phase factors in the geometrically nonequivalent IP and OPs, and the detailed structure of the phase diagram of the multilayered cuprates constitutes an important problem for study. However, all these issues have not yet been fully investigated so far in the literature.

In this Letter, we study the microscopic coexistence and interplay between HTSC and AFM orders within a proximity scenario. An exotic intramultilayer AFM scattering term is proposed to elucidate the appreciable enlargement of the range for the mixed phase. Besides, once isolated CuO$_{2}$ planes coupled together by interlayer tunneling processes, originally degenerate states will be broken down to certain peculiar configurations, which allows us to identify the significance of such layer arrangements of SC and AFM phases in settling the ground state of multilayered cuprates.

{\em Multilayered $t$-$J$ model and Gutzwiller projection.}---In the present paper we shall focus mainly on the system of trilayer HTSC copper oxides and treat it as a prototype to probe the SC and AFM properties of multilayer cuprates. For simplicity, the possibilities of interlayer electron pairing and direct interactions between two OPs are ignored, and we only consider intralayer Cooper pairing and adjacent interlayer couplings within one unit cell of a trilayer cuprate. $t$-$J$ model is adopted to describe the microscopic electronic structure of each layer. The effective free energy function describing such trilayer systems can be written as
\begin{eqnarray}
& &\mathcal{F}=\!\!\!\sum_{\alpha<\beta,\sigma}\!\!\left\{\!-4 g_{t,\alpha} t \chi_{\alpha}\! -\! \left(\!\frac{1}{2} g^{XY}_{s,\alpha} J\!+\!\frac{1}{4} g^{Z}_{s,\alpha} J \!\right)\!\! \left( \Delta^2_{\alpha}\!+\! \chi^2_{\alpha} \right) \right.  \nonumber \\
& &~\left.\!-\!2 g^{Z}_{s,\alpha}J m^2_{\alpha}\!\!-\!\!2\kappa g^{\alpha \beta}_{\sigma,\perp}\! \Gamma^{\sigma}_{\alpha \beta}\!\!-\!\!4\gamma \Delta_{\alpha}  \Delta_{\beta}\!\!+\!U_{\!\perp} \! \delta_{\alpha} \delta_{\beta}\!\!-\!\!\mu_{\alpha} n_{\alpha}\!\right\} 
\label{eq:nof1}
\end{eqnarray}
per site, where three CuO$_{2}$ planes (layer-index $\alpha,\beta\!\!=\!\!1, 2, 3$) share the same parameters of $t$-$J$ model and the same interlayer Coulomb repulsion $U_{\!\perp}$ \cite{ribeiro}, which could redistribute the doped holes over individual layers. A phenomenological pair tunneling Hamiltonian originating from a Josephson-like pair tunneling process is included in $\mathcal{F}$ \cite{zhou}. Two nearest-neighbor electrons with opposite spins are considered to condense into a Cooper pair tunneling in the real space. In Equation (\ref{eq:nof1}), we define such singlet pairing amplitudes as $\Delta_{\alpha}\! \equiv \! \langle c^{(\alpha)}_{i,\uparrow} c^{(\alpha)}_{j,\downarrow} - c^{(\alpha)}_{i,\downarrow} c^{(\alpha)}_{j,\uparrow} \rangle_0$. The notation $\langle \cdots \rangle_{0}$ indicates the expectation value in the unprojected state. The other two order parameters decoupled from the spin-spin superexchange interaction are defined as $\chi_{\alpha}\! \equiv \!\langle c^{(\alpha)\dag}_{i,\uparrow} c^{(\alpha)}_{j,\uparrow}\! +\! c^{(\alpha)\dag}_{i,\downarrow} c^{(\alpha)}_{j,\downarrow} \rangle_{0}$ and $m_{\alpha}\! \equiv \!e^{i\mathbf{Q}\mathbf{R}_{i}} \langle \text{S}^{Z}_{i,\alpha} \rangle_{0}$. $\Gamma^{\sigma}_{\alpha \beta}$ represents a spin-oriented magnetic scattering between adjacent layers deduced from the channel of single electron tunneling, whose form is as follows \cite{supplemetnal}:
\begin{equation}
\Gamma^{\sigma}_{\alpha \beta}\!=\!\frac{1}{N}\sum_{\mathbf{k}} \eta^2_{\mathbf{k}} \left\langle c^{(\alpha)\dagger}_{\mathbf{k},\sigma} c^{(\beta)}_{\mathbf{k+Q},\sigma} \!-\! c^{(\alpha)\dagger}_{\mathbf{k+Q},\sigma} c^{(\beta)}_{\mathbf{k},\sigma}\!+\! \textrm{H.c.}\right\rangle_0\!.
\end{equation}
The summation of $\mathbf{k}$ runs over the reduced Brillouin zone: $-\pi \!<\! k_{x} \pm k_{y} \!\leqslant\! \pi$, the nesting vector $\mathbf{Q}\! \equiv \!(\pi,\pi)$, and $\eta_{\mathbf{k}}\!=\!\cos k_x \!-\! \cos k_y$ is a $d$-wave symmetrical factor \cite{chakravarty2}. Here $\kappa$ and $\gamma$ are strengths of interlayer AFM scattering and Cooper pair tunneling. By deploying extended Gutzwiller approximation \cite{fczhang,ogata} to relax the constraint of no doubly occupied sites, we substitute the projected order parameters with the above-defined mean fields in the unprojected Hilbert space multiplied by various statistical counting factors, which results in the free energy function $\mathcal{F}$. The detailed analytic forms of these Gutzwiller factors can be found in Ref. \cite{ogata} and a simplified version for the renormalized factor $g^{\alpha \beta}_{\sigma,\perp}$ is effectively expressed in terms of the doping levels and AFM moments of IP and OPs \cite{supplemetnal},
\begin{eqnarray}
& &g^{\alpha \beta}_{\sigma,\perp}=\sqrt{\frac{2\delta_{\alpha} (1\!-\!\delta_{\alpha})}{1\!-\!\delta^2_{\alpha}\!+\!4 m^2_{\alpha}} \frac{2\delta_{\beta} (1\!-\!\delta_{\beta})}{1\!-\!\delta^2_{\beta}\!+\!4 m^2_{\beta}}} \nonumber \\
& & \times\! \left(\!\! \sqrt{\frac{1\!+\!\delta_{\alpha}\!-\!2\sigma m_{\alpha}}{1\!+\!\delta_{\alpha}\!+\!2\sigma m_{\alpha}} \frac{1\!+\!\delta_{\beta}\!+\!2\sigma m_{\beta}}{1\!+\!\delta_{\beta}\!-\!2\sigma m_{\beta}}}\!-\!(\sigma\!\rightarrow\! -\sigma) \!\!\right)\!\!.
\end{eqnarray}
This novel interlayer scattering between different magnetic Brillouin zones mimics a hidden AFM correlation in the intramultilayer magnetic structure, and it produces a desired consequence of enhancing AFM moments with the HTSC intact. Apparently, such kinds of interlayer magnetic couplings are solely arising from the strong electron correlation effects in underdoped cuprates and once the AFM moments vanish, this channel will be switched off owing to the shrinking of $g^{\alpha \beta}_{\sigma,\perp}$. Throughout the whole numerical calculation, we set $t\!=\!300$ meV as the energy unit and $J\!=\!0.3t$, which are typical parameters in the $t$-$J$ model.

{\em Enlarged coexistent region of AFM and HTSC.}---Following standard variational procedure, we minimize the free energy $\mathcal{F}$ to obtain varied order parameters from a set of self-consistency equations. The stable local energy minima can be found within restricted parameter ranges by iteration method. In the presence of commensurate AFM order, the Brillouin zone is fold into two parts, which enables us to construct intra- and interlayer magnetic scatterings with nesting vector $\mathbf{Q}$. The fully entangled Gutzwiller renormalization factors further generate strongly nonlinear effects on the self-consistently determined mean-field order parameters, and the explicit dependence of $g^{\alpha \beta}_{\sigma,\perp}$ on the spin indices and AFM moments from nonequivalent IP and OPs plays another essential role to lift the degeneracy of intramultilayer magnetic structures.

\begin{figure}
\begin{center}
\includegraphics[height=6.3cm]{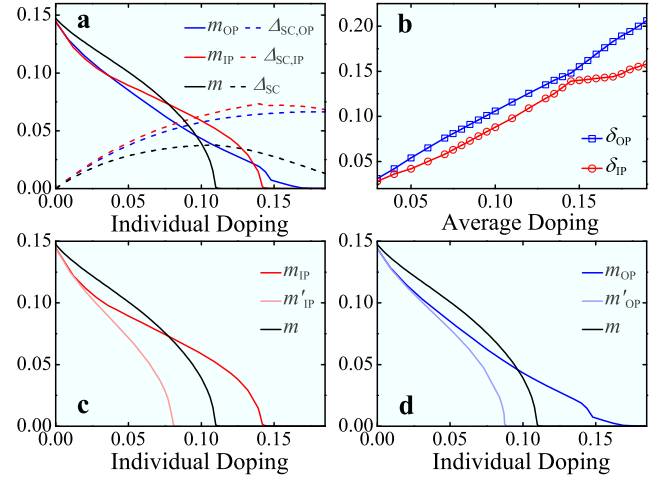}
\caption{\label{fig:fig1} (color online). The staggered AFM moments $m_{\alpha}$ and the projected SC order parameters $\Delta_{\textrm{SC},\alpha}$ in the IP and OPs for the trilayer system as functions of individual doping levels $\delta_{\alpha}$, with $\kappa\!\!=\!\!0.23t,~\gamma\!\!=\!\!0.13t,~U_{\!\!\perp}\!\!=\!\!1.6t$. Corresponding orderings $m$ and $\Delta_{\textrm{SC}}$ for the single-layer $t$-$J$ model were shown for comparison $\mbox{[panel }\textbf{a}\mbox{]}$. The carrier distributions in IP and OPs are plotted as functions of average doping $\delta$ [panel $\textbf{b}$]. The calculated AFM moment versus doping in each layer with the chosen parameters was compared with cases of $\kappa\!\!=\!\!\gamma\!\!=\!\!U_{\!\!\perp}\!\!=\!\!0$ ($m$, black curves) and $\kappa\!\!=\!\!0,~\gamma\!\!=\!\!0.13t,~U_{\!\!\perp}\!\!=\!\!1.6t$ ($m'_{\alpha}$, light red and blue curves) for IP [panel $\textbf{c}$] and OPs [panel $\textbf{d}$].}
\end{center}
\end{figure}

In Fig.~$\ref{fig:fig1}$a, we plot the staggered magnetizations $m_{\alpha}$ and the projected SC order parameters $\Delta_{\textrm{SC},\alpha} \!\equiv \! g_{\Delta} \Delta_{\alpha}$ in both IP (red curves) and OP (blue curves) as functions of individual dopings $\delta_{\alpha}$. The solid and dash lines represent AFM moments and projected SC gaps, respectively. For comparison, we have used the black solid and dash lines to denote the results of decoupled single-layer $t$-$J$ model. A critical doping $\delta_{c}$ can be resolved around $0.1$ in the black solid line, below which AFM order coexists uniformly with SC order \cite{ogata}. As the values of parameters increasing to $\kappa\!\!=\!\!0.23t$, $\gamma\!\!=\!\!0.13t$ and $U_{\!\perp}\!\!=\!\!1.6t$, the ground-state layer configuration of the trilayer system in the mixed regime favors an intramultilayer AFM arrangement with coherent superconductivity. The staggered magnetic phases $m_{\alpha}$ also become much stiffer against dopings for both inner and outer layers. The resulting quantum critical doping points $\delta_{c,\alpha}$ beyond which $m_{\alpha}$ diminish steeply are $0.14$ and $0.17$ for the IP and OPs, respectively. Our result qualitatively agrees with the NMR observation of trilayer cuprates $\textrm{Ba}_{2}\textrm{Ca}_{2}\textrm{Cu}_{3}\textrm{O}_{6}(\textrm{F,O})_{2}$ \cite{shimizu2}. This increase of $\delta_{c,\alpha}$ to higher magnitudes arises from interlayer AFM scatterings and the high entanglement of order parameters in the limit of strong electron correlation. The interlayer quantum tunneling of Cooper pairs further immensely promotes the SC gaps throughout the whole system, and the maximum value of the projected SC order parameter in the IP is nearly twice larger than that of the single-layer cuprate, which is consistent with the enhanced superconducting gap observed by the ARPES experiment in Bi$2223$ \cite{ideta}. Moreover, around $\delta_{c,\alpha}$, the SC orders for both IP and OP always reach their maxima.

Another subtle phenomenon in realistic multilayered materials is the nonhomogeneous doping distribution among inner and outer cuprate planes \cite{mukuda,shimizu1,shimizu2,ideta,kotegawa}. We describe such an essential effect by $U_{\!\perp}$ \cite{ribeiro}. Considering the doped holes have a gradually diffusive process from outer planes to the inner, two different values have further been assigned to $\{\mu_{\alpha}\}$ to exemplify a possible layer gradient of chemical potential. The self-consistently calculated charge distribution is shown in Fig.~$\ref{fig:fig1}$b with the same values of parameters as in panel a. We find that throughout the average doping axis, the doping level of OP is always greater than that for the IP and their imbalanced value is about $0.05$ when the averaged doping $\delta$ equals $0.19$. This is consistent with the experimental observations \cite{kotegawa}. Furthermore, the concentrations of doped holes in the inner and outer planes are both monotonically increasing functions of $\delta$, except that there exist kinks at the critical average doping $0.15$, beyond which the system is purely superconducting. Finally, the coexistence of AFM and HTSC might slightly shrink the resulting charge imbalance between IP and OP.

Due to the competition between emergent AFM and enhanced HTSC, the values of AFM moments of the trilayer system would be greatly suppressed in the region close to $\delta_{c}$, if the scattering strength $\kappa$ was reduced to zero. However, once $\kappa$ becomes comparable to $\gamma$, this suppression can be reversely changed, especially in the regime with large doping concentrations, as demonstrated in Fig.~$\ref{fig:fig1}$c and d, where we compare the results of staggered magnetization for cases with sets of parameters: $\kappa\!\!=\!\!0.23t,~\gamma\!\!=\!\!0.13t,~U_{\!\perp}\!\!=\!\!1.6t$ and $\kappa\!\!=\!\!0,~\gamma\!\!=\!\!0.13t,~U_{\!\perp}\!\!=\!\!1.6t$, as well as $\kappa\!\!=\!\!\gamma\!\!=\!\!U_{\!\perp}\!\!=\!\!0$ in both IP and OP. One of the unusual features of the designed interlayer AFM coupling is its compatibility with the interlayer Cooper pair tunneling, which means we can conversely compensate the suppression of AFM moments via magnetic scattering without reducing the enhanced superconductivity from the channel of pair tunneling. This unique combination and interplay between these two channels underpins the realization of the extended coexistence of AFM and HTSC in our trilayer $t$-$J$ model and the achieved charge imbalance further facilitates this phenomenon.

{\em Predicted phase diagrams of trilayer cuprates.}---We would like now to analyse the relative layer configurations of coexisting AFM and SC order parameters in the presence of both interlayer magnetic scattering and Cooper pair tunneling. Since these two orderings have the Ising symmetry, the interlayer couplings will drive four distinct stacking patterns between IP and OP in one unit cell when only thermodynamic stable phases are considered. A candidate state here is specified by its phase arrangements of both AFM and SC orders among three layers. We distinguish the layer arrangements of AFM moments into two categories: the ferro-type structure, namely [$\mathop{\rm sgn}(\!\frac{m_{\rm{IP}}}{m_{\rm{OP}}}\!)\!\!=\!\!1$], which indicates that the interlayer configuration of AFM moments is ferromagnetically aligned in one unit cell, and the antiferro-type structure, namely [$\mathop{\rm sgn}(\!\frac{m_{\rm{IP}}}{m_{\rm{OP}}}\!)\!\!=\!\!-1$], denoting the AFM arrangement. In the same way, we have inphase [$\mathop{\rm sgn}(\!\frac{\Delta_{\rm{IP}}}{\Delta_{\rm{OP}}}\!)\!\!=\!\!1$] and out-of-phase [$\mathop{\rm sgn}(\!\frac{\Delta_{\rm{IP}}}{\Delta_{\rm{OP}}}\!)\!\!=\!\!-1$] SC states. Therefore, totally there exist four possible degenerate ground states with distinct symmetries in layer configurations of order parameters when three layers are decoupled ($\kappa\!\!=\!\!\gamma\!\!=\!\!0$). However, interlayer tunnelings, as weak perturbations, will spoil originally degenerate equilibrium state to some special symmetry broken configuration. Possible abrupt phase transitions between these correlated states will be uncloaked through tuning the strengths of $\kappa$ and $\gamma$ and/or the averaged doping $\delta$. In the following, we will use $(\pm,\pm)$ to denote these four states individually and the first and second \lq\lq $\pm$''s represent the intramultilayer arrangements of AFM and SC orderings, respectively. The first \lq\lq $+$'' sign on the left refers to the ferro-type structure for the AFM order and the second \lq\lq $+$'' sign on the right refers to the inphase state for the SC order, while the \lq\lq $-$'' signs are adopted for the opposite magnetic and superconducting configurations.

\begin{figure}
\begin{center}
\includegraphics[height=6.3cm]{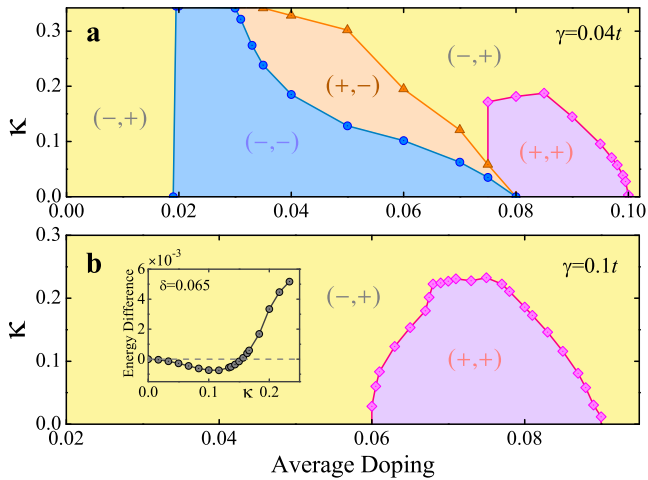}
\caption{\label{fig:fig2} (color online). The ground-state phase diagram for the triple-layer system in the weak Josephson-coupling limit with $\gamma\!\!=\!\!0.04t$ [panel $\textbf{a}$], as well as the phase diagram in the strong limit of Josephson coupling with $\gamma\!\!=\!\!0.1t$ [panel $\textbf{b}$]. The inset of panel $\textbf{b}$ depicts the free energy difference between states $(+,+)$ and $(-,+)$ when tuning $\kappa$ at $\delta\!\!=\!\!0.065$. See text for the definitions of states $(\pm,\pm)$.}
\end{center}
\end{figure}

Minimization of the free energy $\mathcal{F}$ in the order parametric space leads to the genuine ground state with specified configurational symmetries. The internal competition, entanglement and renormalization of orderings further combine together to manifest the essential interplay between AFM and HTSC in multilayer materials. Physically, we know that sufficiently large interlayer magnetic scattering will drive the triple-layer system to exhibit the antiferro-type magnetic structure, while the strong Cooper pair tunneling will give rise to the inphase SC state. Nevertheless, the detailed evolution, when sweeping through the relevant parametric space $(\kappa,\gamma,\delta)$, is still not clear for multilayered $t$-$J$ model. In view of the qualitatively consistent results derived above from the trilayer proximity model, in the following, we will address this crucial problem based on analysing the same free energy function $\mathcal{F}$.

We first consider the weak Josephson-coupling limit, from which we can extract some generic features about the transitions between these correlated ground-state configurations. Figure~$\ref{fig:fig2}$a is a phase diagram for the AFM and HTSC coexisting region plotted in terms of $\kappa$ and $\delta$ with $\gamma\!=\!0.04t$ and $U_{\!\perp}$ fixed to be $1.6t$. Four possible layer configurations are all present in different parts of the $\kappa$-$\delta$ plane. In the almost undoped regime, the antiferro-inphase state $(-,+)$ always possesses the lowest free energy than other arrangements along the $\kappa$ axis. When gradually increasing $\delta$ to the intermediate underdoped regime, the antiphase SC state emerges as the ground state with a variety of magnetic structures: one first enters a region in which the symmetry $(-,-)$ dominates, and then with increased $\kappa$, state $(+,-)$ supersedes $(-,-)$ to be energy favored. As further lifting $\kappa$ above a critical value, the system will undergo an abrupt phase transition accompanying with both SC and AFM phase-changes: $(+,-)\!\Leftrightarrow\!(-,+)$. These entangled simultaneous phase-shifts offer a unique manifestation of the interplay and correlation between the symmetries of AFM and SC phase arrangements, and the sufficient renormalization of chemical potential, the resulting nonhomogeneous charge distribution and the energy competition between AFM, SC, and kinetic orderings are the major underlying driving force of this transition. Experimental route to probe the signature of such a phase transition might include using phase-sensitive facilities \cite{cctsuei}. As $\delta$ further approaching $\delta_{c}$, the intramultilayer SC phases in one unit cell will incline to form a coherent arrangement with a phase transition from the ferro-type magnetic structure to the antiferro-type when continuously promoting the strength of $\kappa$. Beyond $\delta_{c}$, the AFM ordering of the system is mainly sustained by the channel of interlayer magnetic scattering, thus the only allowed magnetic structure is of the antiferro-type.

In the strong limit of interlayer Josephson coupling, the out-of-phase SC state will become unstable and eventually fade out for all the doping concentrations we consider. The inphase SC state prevails to be the ground state with various magnetic structures when $\gamma\!=\!0.1t$. We find that if the averaged doping is within intermediate underdoped regime $0.06\!\!<\!\!\delta\!\!<\!\!0.09$, the ferro-inphase state $(+,+)$ will have a lower free energy than the antiferro-inphase state $(-,+)$ when $\kappa$ is varying between $0t$ to $0.2t$. The resulting phase diagram is presented in Figure~$\ref{fig:fig2}$b, where we calculate a dome-like phase boundary separating these two phases. To gain a concrete understanding of this magnetic phase transition, we plot the free energy difference between states $(+,+)$ and $(-,+)$ as a function of $\kappa$ in the inset of Figure~$\ref{fig:fig2}$b. The average doping is chosen to be $0.065$ and the energy difference is defined as $\Delta \mathcal{F}\!=\!\mathcal{F}_{(+,+)}\!-\!\mathcal{F}_{(-,+)}$. The boundary between these two phases can be obtained by nontrivial values of $\kappa$ and $\delta$ satisfying $\Delta \mathcal{F}\!=\!0$. In the other region of the phase diagram, the ground state of the trilayer system should have the $(-,+)$ symmetry and exhibit the antiferro-type magnetic structure with enhanced superconductivity.

While our minimal model (\ref{eq:nof1}) is simple, implications of the AFM-SC mixed phase and the mapped phase diagrams can be readily generalized to more complex layer materials. Specifically, if we increase $n$ from $3$ to $4$ and to $5$, the system might turn out to decay from a fully \textit{inphase} SC state to the partially \textit{out-of-phase} SC state with narrower pairing gaps, thus giving rise to a lower $\textit{T}_\textit{c}$. In addition, with more layers stacking together in one building block of multilayered cuprate, the layer gradient of chemical potential will be amplified in the influence of interlayer magnetic scattering, and we need to dope more holes into the system to completely suppress the AFM order in the innermost CuO$_{2}$ plane, which reversely indicates the microscopic coexistence of AFM with HTSC can survive to a larger average doping.

{\em Conclusion.}---In summary, we have performed a systematic investigation of the interlayer coupling effects on trilayer cuprates, which brings into full focus the issues concerning novel intramultilayer AFM scatterings in producing the observed broadening of AFM and HTSC coexisting region and abrupt correlated phase transitions between four different AFM-SC states for the trilayer system. Our predicted phase diagrams might serve as the first attempt to clarify symmetry-sensitive structures in layer materials and also to highlight the profound consequences of interlayer phase arrangements in triggering the unconventional interplay between HTSC and AFM in multilayered high-$\textit{T}_\textit{c}$ superconductors.

This work was supported by the Summer Research Fellowship at the University of Minnesota (C.C.), the A$3$ Foresight Program (C.C., A.F., and Y.C.), the State Key Programs of China (Grant Nos. $2009$CB$929204$ and $2012$CB$921604$), the National Natural Science Foundation of China (Grant Nos. $10874032$ and $11074043$) and Shanghai Municipal Government (Y.C.), and the Robert A. Welch Foundation (C.S.T.) under the Grant No. E-$1146$.

\end{document}


\title{Supplemental Material for Coexistence of Enhanced Superconductivity and Antiferromagnetism: Possible Correlated Phase Transitions in Trilayer High-$\textit{T}_\textit{c}$ Cuprates}

\author{Chun Chen}
\email[Corresponding author: ]{cchen@physics.umn.edu}
\affiliation{School of Physics and Astronomy, University of Minnesota, Minneapolis, Minnesota 55455, USA}

\author{A. Fujimori}
\affiliation{Department of Physics, University of Tokyo, Bunkyo-ku, Tokyo 113-0033, Japan}

\author{C. S. Ting}
\affiliation{Texas Center for Superconductivity and Department of Physics, University of Houston, Houston, Texas 77204, USA}

\author{Yan Chen}
\email[]{yanchen99@fudan.edu.cn}
\affiliation{Department of Physics, Surface Physics Laboratory (National Key Laboratory) and Laboratory of Advanced Materials, Fudan University, Shanghai 200433, China}

\date{\today}

\maketitle

\onecolumngrid

\section{Derivations of the intramultilayer antiferromagnetic scattering amplitude $\Gamma^{\sigma}_{\alpha \beta}$ and \\ the Gutzwiller renormalization factor $g^{\alpha \beta}_{\sigma,\perp}$}

In this supplemental material, we will first provide the motivation for introducing the interlayer antiferromagnetic scattering and then give out detailed deductions for the explicit forms of $\Gamma^{\sigma}_{\alpha \beta}$ and $g^{\alpha \beta}_{\sigma,\perp}$.

To avoid possible confusions, we need to define and fix two conventions when dealing with antiferromagnetism in the extended Gutzwiller approximation of a single-layer $t$-$J$ model in a square lattice containing $N$ sites:
\begin{equation}
\tilde{\mathcal{H}}_{t\mbox{-}J}= -t \sum_{\langle i,j \rangle, \sigma} (\tilde{c}^{\dagger}_{i\sigma} \tilde{c}_{j\sigma}+\textrm{H.c.})+J\sum_{\langle i,j\rangle} \tilde{\textbf{S}}_{i} \cdot \tilde{\textbf{S}}_{j}-\mu \sum_{\alpha, i}\tilde{n}_{i},
\end{equation}
where $\tilde{c}^{\dagger}$ and $\tilde{c}$ represent the electron creation and annihilation operators in the projected Hilbert space. In the presence of antiferromagnetism, we shall divide the square lattice into two sublattices: sites $A$ with net spin up and sites $B$ with net spin down. Under the Gutzwiller projection, we substitute each operator $\tilde{c}^{\dagger}_{i\sigma}$ ($\tilde{c}_{i\sigma}$) in the hopping terms with a corresponding unprojected operator $c^{\dagger}_{i\sigma}$ ($c_{i\sigma}$) multiplied by a classical Gutzwiller factor $g^{t}_{i \sigma}$. For clarity, we demonstrate the first convention we use in defining a simplified version of $g^{t}_{i \sigma}$ as follows:
\begin{equation}
g^{t}_{i \sigma}=\left\{
\begin{array}{cl}
\sqrt{\frac{2\delta(1-\delta)}{1-\delta^{2}+4 m^{2}} \frac{1+\delta+\sigma 2 m}{1+\delta-\sigma 2 m}} & \mbox{if}~~i=A \\
\sqrt{\frac{2\delta(1-\delta)}{1-\delta^{2}+4 m^{2}} \frac{1+\delta-\sigma 2 m}{1+\delta+\sigma 2 m}} & \mbox{if}~~i=B
\end{array}\right. ,
\label{eq:gt}
\end{equation}
where $\delta$ is the doping concentration and $m\! \equiv \!e^{i\mathbf{Q}\mathbf{R}_{i}} \langle \text{S}^{Z}_{i} \rangle_{0}$ is the staggered magnetic moment in the unprojected state \cite{wqchen}. When Fourier transforming the Hamiltonian from one space to the other, we adopt the following correspondence between operators $c_{A\mathbf{k}\sigma}$, $c_{B\mathbf{k}\sigma}$ and $c_{\mathbf{k},\sigma}$, $c_{\mathbf{k+Q},\sigma}$ \cite{qsyuan}:
\begin{equation}
c_{\mathbf{k},\sigma}=\frac{c_{A\mathbf{k}\sigma}+c_{B\mathbf{k}\sigma}}{\sqrt{2}},~~~c_{\mathbf{k+Q},\sigma}=\frac{c_{A\mathbf{k}\sigma}-c_{B\mathbf{k}\sigma}}{\sqrt{2}}.
\end{equation}

Armed with these conventions, we now illustrate straightforwardly that for a multilayered $t$-$J$ model, the proposed amplitude of the spin-oriented magnetic scattering $\Gamma^{\sigma}_{\alpha \beta}$ can be regarded as an effective order parameter of the intramultilayer magnetic structure. Temporarily, we neglect the $d$-wave factor $\eta^2_{\mathbf{k}}$ and redefine $\Gamma^{\sigma}_{\alpha \beta}$ as
\begin{equation}
\bar{\Gamma}^{\sigma}_{\alpha \beta}\!=\!\frac{1}{N}\sum_{\mathbf{k}} \left\langle c^{(\alpha)\dagger}_{\mathbf{k},\sigma} c^{(\beta)}_{\mathbf{k+Q},\sigma} \!-\! c^{(\alpha)\dagger}_{\mathbf{k+Q},\sigma} c^{(\beta)}_{\mathbf{k},\sigma}\!+\! \textrm{H.c.}\right\rangle_0\!,
\end{equation}
where the summation of $\mathbf{k}$ is summed over the magnetic Brillouin zone. It is easy to show that if the AFM moments of IP ($\alpha$) and OP ($\beta$) are aligned ferromagnetically in one unit cell, namely $\mathop{\rm sgn}(\frac{m_{\alpha}}{m_{\beta}})\!=\!1$, the value of $\bar{\Gamma}^{\sigma}_{\alpha \beta}$ is close to zero:
\begin{eqnarray}
\bar{\Gamma}^{\sigma}_{\alpha \beta}\!&=&\!\frac{1}{N}\sum_{\mathbf{k}} \left\langle c^{(\alpha)\dagger}_{\mathbf{k},\sigma} c^{(\beta)}_{\mathbf{k+Q},\sigma} \!-\! c^{(\alpha)\dagger}_{\mathbf{k+Q},\sigma} c^{(\beta)}_{\mathbf{k},\sigma}\!+\! \textrm{H.c.}\right\rangle_0\! \nonumber \\
&=&\frac{1}{N}\sum_{\mathbf{k}} \left\langle \left(\frac{c^{(\alpha)\dagger}_{A\mathbf{k}\sigma}+c^{(\alpha)\dagger}_{B\mathbf{k}\sigma}}{\sqrt{2}}\right) \!\! \left(\frac{c^{(\beta)}_{A\mathbf{k}\sigma}-c^{(\beta)}_{B\mathbf{k}\sigma}}{\sqrt{2}}\right)\!-\!\left(\frac{c^{(\alpha)\dagger}_{A\mathbf{k}\sigma}-c^{(\alpha)\dagger}_{B\mathbf{k}\sigma}}{\sqrt{2}}\right) \!\! \left(\frac{c^{(\beta)}_{A\mathbf{k}\sigma}+c^{(\beta)}_{B\mathbf{k}\sigma}}{\sqrt{2}}\right) \!+\! \textrm{H.c.} \right\rangle_0\! \nonumber \\
&=&\frac{1}{N}\sum_{\mathbf{k}} \left\langle c^{(\alpha)\dagger}_{B\mathbf{k}\sigma} c^{(\beta)}_{A\mathbf{k}\sigma}\!-\!c^{(\alpha)\dagger}_{A\mathbf{k}\sigma} c^{(\beta)}_{B\mathbf{k}\sigma} \!+\! \textrm{H.c.} \right\rangle_0\! \nonumber \\
&\simeq&0.
\end{eqnarray}
On the contrary, when the staggered AFM moments are inclined to be antiferromagnetically arranged in a unit cell, namely $\mathop{\rm sgn}(\frac{m_{\alpha}}{m_{\beta}})\!=\!-1$, the value of $\bar{\Gamma}^{\sigma}_{\alpha \beta}$ is approaching a finite number:
\begin{eqnarray}
\bar{\Gamma}^{\sigma}_{\alpha \beta}\!&=&\!\frac{1}{N}\sum_{\mathbf{k}} \left\langle c^{(\alpha)\dagger}_{\mathbf{k},\sigma} c^{(\beta)}_{\mathbf{k+Q},\sigma} \!-\! c^{(\alpha)\dagger}_{\mathbf{k+Q},\sigma} c^{(\beta)}_{\mathbf{k},\sigma}\!+\! \textrm{H.c.}\right\rangle_0\! \nonumber \\
&=&\frac{1}{N}\sum_{\mathbf{k}} \left\langle \left(\frac{c^{(\alpha)\dagger}_{A\mathbf{k}\sigma}+c^{(\alpha)\dagger}_{B\mathbf{k}\sigma}}{\sqrt{2}}\right) \!\! \left(\frac{c^{(\beta)}_{B\mathbf{k}\sigma}-c^{(\beta)}_{A\mathbf{k}\sigma}}{\sqrt{2}}\right)\!-\!\left(\frac{c^{(\alpha)\dagger}_{A\mathbf{k}\sigma}-c^{(\alpha)\dagger}_{B\mathbf{k}\sigma}}{\sqrt{2}}\right) \!\! \left(\frac{c^{(\beta)}_{B\mathbf{k}\sigma}+c^{(\beta)}_{A\mathbf{k}\sigma}}{\sqrt{2}}\right) \!+\! \textrm{H.c.} \right\rangle_0\! \nonumber \\
&=&\frac{1}{N}\sum_{\mathbf{k}} \left\langle c^{(\alpha)\dagger}_{B\mathbf{k}\sigma} c^{(\beta)}_{B\mathbf{k}\sigma}\!-\!c^{(\alpha)\dagger}_{A\mathbf{k}\sigma} c^{(\beta)}_{A\mathbf{k}\sigma} \!+\! \textrm{H.c.} \right\rangle_0\! \nonumber \\
&=&\frac{1}{N}\left\langle\left( \sum_{B} c^{(\alpha)\dagger}_{B\sigma} c^{(\beta)}_{B\sigma}\!+\! \textrm{H.c.}\right)\!-\!\left(\sum_{A} c^{(\alpha)\dagger}_{A\sigma} c^{(\beta)}_{A\sigma}\!+\! \textrm{H.c.} \right)\right\rangle_0\neq0,
\end{eqnarray}
owing to the emergence of antiferromagnetic ordering and the Pauli exclusion principle. After averaging over the $d$-wave symmetrical factor $\eta^2_{\mathbf{k}}$, we find that under the same set of parameters, numerically the value of $\Gamma^{\sigma}_{\alpha \beta}$ calculated in the antiferro-type magnetic structure is one order greater than the value of $\Gamma^{\sigma}_{\alpha \beta}$ evaluated in the ferro-type magnetic structure. Motivated by the above observation, we dub $\Gamma^{\sigma}_{\alpha \beta}$ and the related process a sort of intramultilayer antiferromagnetic scattering (coupling).

With the identification of $\Gamma^{\sigma}_{\alpha \beta}$ as the amplitude of an interlayer magnetic scattering, we are going to construct an across-the-interface interaction which has an analogous analytic form like $\Gamma^{\sigma}_{\alpha \beta}$. The methodology we deploy here is instead of direct model building, we choose to filter out such kinds of magnetic interactions from the various channels of interlayer single electron tunneling. Within the filtering process, we will find the explicit form of the Gutzwiller renormalized factor $g^{\alpha\beta}_{\sigma,\perp}$. As an illustration, we first consider the onsite single electron tunneling between adjacent IP ($\alpha$) and OP ($\beta$) in one unit cell of a trilayer cuprate which is exhibiting the antiferro-type magnetic structure:
\begin{equation}
\tilde{\mathcal{H}}_{\perp}=-t_{\perp} \sum_{A} \left(\tilde{c}^{(\alpha)\dagger}_{A \uparrow}\tilde{c}^{(\beta)}_{A \uparrow}+\tilde{c}^{(\beta)\dagger}_{A \uparrow}\tilde{c}^{(\alpha)}_{A \uparrow}+\tilde{c}^{(\alpha)\dagger}_{A \downarrow}\tilde{c}^{(\beta)}_{A \downarrow}+\tilde{c}^{(\beta)\dagger}_{A \downarrow}\tilde{c}^{(\alpha)}_{A \downarrow}\right)-t_{\perp} \sum_{B} \left(\tilde{c}^{(\alpha)\dagger}_{B \uparrow}\tilde{c}^{(\beta)}_{B \uparrow}+\tilde{c}^{(\beta)\dagger}_{B \uparrow}\tilde{c}^{(\alpha)}_{B \uparrow}+\tilde{c}^{(\alpha)\dagger}_{B \downarrow}\tilde{c}^{(\beta)}_{B \downarrow}+\tilde{c}^{(\beta)\dagger}_{B \downarrow}\tilde{c}^{(\alpha)}_{B \downarrow}\right).
\end{equation}
Under the extended Gutzwiller approximation, the renormalized Hamiltonian of $\tilde{\mathcal{H}}_{\perp}$ is
\begin{eqnarray}
\mathcal{H}_{\perp}&=&-t_{\perp} \sum_{A} \left(g^{t(\alpha)}_{A \uparrow}g^{t(\beta)}_{A \uparrow}c^{(\alpha)\dagger}_{A \uparrow}c^{(\beta)}_{A \uparrow}+g^{t(\alpha)}_{A \uparrow}g^{t(\beta)}_{A \uparrow}c^{(\beta)\dagger}_{A \uparrow}c^{(\alpha)}_{A \uparrow}+g^{t(\alpha)}_{A \downarrow}g^{t(\beta)}_{A \downarrow}c^{(\alpha)\dagger}_{A \downarrow}c^{(\beta)}_{A \downarrow}+g^{t(\alpha)}_{A \downarrow}g^{t(\beta)}_{A \downarrow}c^{(\beta)\dagger}_{A \downarrow}c^{(\alpha)}_{A \downarrow}\right) \nonumber \\
&&-t_{\perp} \sum_{B} \left(g^{t(\alpha)}_{B \uparrow}g^{t(\beta)}_{B \uparrow}c^{(\alpha)\dagger}_{B \uparrow}c^{(\beta)}_{B \uparrow}+g^{t(\alpha)}_{B \uparrow}g^{t(\beta)}_{B \uparrow}c^{(\beta)\dagger}_{B \uparrow}c^{(\alpha)}_{B \uparrow}+g^{t(\alpha)}_{B \downarrow}g^{t(\beta)}_{B \downarrow}c^{(\alpha)\dagger}_{B \downarrow}c^{(\beta)}_{B \downarrow}+g^{t(\alpha)}_{B \downarrow}g^{t(\beta)}_{B \downarrow}c^{(\beta)\dagger}_{B \downarrow}c^{(\alpha)}_{B \downarrow}\right),
\end{eqnarray}
where we have adopted the simplified version of Gutzwiller projection factor $g^{t}_{i\sigma}$ defined in Equation (\ref{eq:gt}). Since we have assumed working in an antiferro-type magnetic structure, we shall clearly distinguish $g^{t(\alpha)}_{i\sigma}$ and $g^{t(\beta)}_{i\sigma}$ to be consistent with the convention of Equation (\ref{eq:gt}):
\begin{equation}
g^{t(\alpha)}_{i \sigma}=\left\{
\begin{array}{cl}
\sqrt{\frac{2\delta_{\alpha}(1-\delta_{\alpha})}{1-\delta^{2}_{\alpha}+4 m^{2}_{\alpha}} \frac{1+\delta_{\alpha}+\sigma 2 m_{\alpha}}{1+\delta_{\alpha}-\sigma 2 m_{\alpha}}} & \mbox{if}~~i=A \\ \sqrt{\frac{2\delta_{\alpha}(1-\delta_{\alpha})}{1-\delta^{2}_{\alpha}+4 m^{2}_{\alpha}} \frac{1+\delta_{\alpha}-\sigma 2 m_{\alpha}}{1+\delta_{\alpha}+\sigma 2 m_{\alpha}}} & \mbox{if}~~i=B
\end{array}\right. ,
\label{eq:ga}
\end{equation}
while
\begin{equation}
g^{t(\beta)}_{i \sigma}=\left\{
\begin{array}{cl}
\sqrt{\frac{2\delta_{\beta}(1-\delta_{\beta})}{1-\delta^{2}_{\beta}+4 m^{2}_{\beta}} \frac{1+\delta_{\beta}-\sigma 2 m_{\beta}}{1+\delta_{\beta}+\sigma 2 m_{\beta}}} & \mbox{if}~~i=A \\ \sqrt{\frac{2\delta_{\beta}(1-\delta_{\beta})}{1-\delta^{2}_{\beta}+4 m^{2}_{\beta}} \frac{1+\delta_{\beta}+\sigma 2 m_{\beta}}{1+\delta_{\beta}-\sigma 2 m_{\beta}}} & \mbox{if}~~i=B
\end{array}\right. .
\label{eq:gb}
\end{equation}
Here we have classified the sites with net spin orientation up in the IP ($\alpha$) into the sublattice $A$ and those with net spin orientation down in the IP into the sublattice $B$. By using the Fourier transformations:
\begin{eqnarray}
c_{A\sigma}&=&\sqrt{\frac{2}{N}}\sum_{\mathbf{k}} e^{i\mathbf{k}\mathbf{R}_{A}} c_{A \mathbf{k}\sigma}, \nonumber \\
c_{B\sigma}&=&\sqrt{\frac{2}{N}}\sum_{\mathbf{k}} e^{i\mathbf{k}\mathbf{R}_{B}} c_{B \mathbf{k}\sigma},
\end{eqnarray}
we express $\mathcal{H}_{\perp}$ in the momentum space as
\begin{eqnarray}
\mathcal{H}_{\perp}&=&-\sum_{\mathbf{k}} t_{\perp}\!(\mathbf{k})\! \left(g^{t(\alpha)}_{A \uparrow}g^{t(\beta)}_{A \uparrow}c^{(\alpha)\dagger}_{A \mathbf{k}\uparrow}c^{(\beta)}_{A \mathbf{k} \uparrow}+g^{t(\alpha)}_{A \uparrow}g^{t(\beta)}_{A \uparrow}c^{(\beta)\dagger}_{A \mathbf{k}\uparrow}c^{(\alpha)}_{A \mathbf{k}\uparrow}+g^{t(\alpha)}_{A \downarrow}g^{t(\beta)}_{A \downarrow}c^{(\alpha)\dagger}_{A \mathbf{k}\downarrow}c^{(\beta)}_{A \mathbf{k}\downarrow}+g^{t(\alpha)}_{A \downarrow}g^{t(\beta)}_{A \downarrow}c^{(\beta)\dagger}_{A \mathbf{k} \downarrow}c^{(\alpha)}_{A\mathbf{k} \downarrow}\right) \nonumber \\
&&-\sum_{\mathbf{k}} t_{\perp}\!(\mathbf{k})\! \left(g^{t(\alpha)}_{B \uparrow}g^{t(\beta)}_{B \uparrow}c^{(\alpha)\dagger}_{B \mathbf{k}\uparrow}c^{(\beta)}_{B \mathbf{k} \uparrow}+g^{t(\alpha)}_{B \uparrow}g^{t(\beta)}_{B \uparrow}c^{(\beta)\dagger}_{B \mathbf{k}\uparrow}c^{(\alpha)}_{B \mathbf{k}\uparrow}+g^{t(\alpha)}_{B \downarrow}g^{t(\beta)}_{B \downarrow}c^{(\alpha)\dagger}_{B \mathbf{k}\downarrow}c^{(\beta)}_{B \mathbf{k}\downarrow}+g^{t(\alpha)}_{B \downarrow}g^{t(\beta)}_{B \downarrow}c^{(\beta)\dagger}_{B \mathbf{k} \downarrow}c^{(\alpha)}_{B\mathbf{k} \downarrow}\right),
\end{eqnarray}
where a $d$-wave symmetric factor is introduced $t_{\perp}\!(\mathbf{k})=t_{\perp}(\cos k_x-\cos k_y)^2$ \cite{chakravarty}. However, once we translate the above expression into the $\mathbf{k}$ and $\mathbf{k\!+\!Q}$ formalism, the desired new terms will emerge due to the special form of $g^{t}_{i\sigma}$. According to the following relations:
\begin{eqnarray}
&&c^{(\alpha)}_{A\mathbf{k}\sigma}=\frac{c^{(\alpha)}_{\mathbf{k},\sigma}+c^{(\alpha)}_{\mathbf{k+Q},\sigma}}{\sqrt{2}},~~~
c^{(\alpha)}_{B\mathbf{k}\sigma}=\frac{c^{(\alpha)}_{\mathbf{k},\sigma}-c^{(\alpha)}_{\mathbf{k+Q},\sigma}}{\sqrt{2}}, \nonumber \\
&&c^{(\beta)}_{A\mathbf{k}\sigma}=\frac{c^{(\beta)}_{\mathbf{k},\sigma}-c^{(\beta)}_{\mathbf{k+Q},\sigma}}{\sqrt{2}},~~~
c^{(\beta)}_{B\mathbf{k}\sigma}=\frac{c^{(\beta)}_{\mathbf{k},\sigma}+c^{(\beta)}_{\mathbf{k+Q},\sigma}}{\sqrt{2}},
\end{eqnarray}
finally, we find that the renormalized Hamiltonian $\mathcal{H}_{\perp}$ of the single electron tunneling can be cast in the form
\begin{eqnarray}
\mathcal{H}_{\perp}&=&-\sum_{\mathbf{k}} t_{\perp}\!(\mathbf{k})\! \left(g^{t(\alpha)}_{B\uparrow} g^{t(\beta)}_{B\uparrow} \!-\! g^{t(\alpha)}_{A\uparrow} g^{t(\beta)}_{A\uparrow} \right)\! \left( c^{(\alpha)\dagger}_{\mathbf{k},\uparrow} c^{(\beta)}_{\mathbf{k+Q},\uparrow} \!-\! c^{(\alpha)\dagger}_{\mathbf{k+Q},\uparrow} c^{(\beta)}_{\mathbf{k},\uparrow} \!+\! \textrm{H.c.}\right) \nonumber \\
& &-\sum_{\mathbf{k}} t_{\perp}\!(\mathbf{k})\! \left(g^{t(\alpha)}_{B\downarrow} g^{t(\beta)}_{B\downarrow} \!-\! g^{t(\alpha)}_{A\downarrow} g^{t(\beta)}_{A\downarrow} \right)\! \left( c^{(\alpha)\dagger}_{\mathbf{k},\downarrow} c^{(\beta)}_{\mathbf{k+Q},\downarrow} \!-\! c^{(\alpha)\dagger}_{\mathbf{k+Q},\downarrow} c^{(\beta)}_{\mathbf{k},\downarrow} \!+\! \textrm{H.c.}\right) \nonumber \\
& &-\sum_{\mathbf{k}} t_{\perp}\!(\mathbf{k})\! \left(g^{t(\alpha)}_{B\uparrow} g^{t(\beta)}_{B\uparrow} \!+\! g^{t(\alpha)}_{A\uparrow} g^{t(\beta)}_{A\uparrow} \right)\! \left( c^{(\alpha)\dagger}_{\mathbf{k},\uparrow} c^{(\beta)}_{\mathbf{k},\uparrow} \!-\! c^{(\alpha)\dagger}_{\mathbf{k+Q},\uparrow} c^{(\beta)}_{\mathbf{k+Q},\uparrow} \!+\! \textrm{H.c.}\right) \nonumber \\
& &-\sum_{\mathbf{k}} t_{\perp}\!(\mathbf{k})\! \left(g^{t(\alpha)}_{B\downarrow} g^{t(\beta)}_{B\downarrow} \!+\! g^{t(\alpha)}_{A\downarrow} g^{t(\beta)}_{A\downarrow} \right)\! \left( c^{(\alpha)\dagger}_{\mathbf{k},\downarrow} c^{(\beta)}_{\mathbf{k},\downarrow} \!-\! c^{(\alpha)\dagger}_{\mathbf{k+Q},\downarrow} c^{(\beta)}_{\mathbf{k+Q},\downarrow} \!+\! \textrm{H.c.}\right),
\label{eq:gf}
\end{eqnarray}
where $t_{\perp}\!(\mathbf{k})$ is modified by a factor of one half and $g^{t(\alpha)}_{i\sigma}$, $g^{t(\beta)}_{i\sigma}$ are defined in Equations (\ref{eq:ga}) and (\ref{eq:gb}), respectively. Several comments should be noticed:
\begin{enumerate}
\item[(i)] Owing to the discrepancy of $g^{t}_{i\sigma}$ in different sublattices, novel terms whose forms are identical to that of the amplitude of intramultilayer antiferromagnetic scattering $\Gamma^{\sigma}_{\alpha \beta}$ have been emerging in the first two rows, which explicitly indicates the existence of possible interlayer antiferromagnetic correlations in multiplane cuprates;
\item[(ii)] From Equation (\ref{eq:gf}), we can directly read out the detailed expression for the effective Gutzwiller renormalization factor $g^{\alpha \beta}_{\sigma,\perp}\equiv g^{t(\alpha)}_{B\sigma} g^{t(\beta)}_{B\sigma} \!-\! g^{t(\alpha)}_{A\sigma} g^{t(\beta)}_{A\sigma}$;
\item[(iii)] Because the renormalized factor $g^{t}_{i\sigma}$ is depending on the spin orientation, originally equally-weighted channels of spin up and spin down are now showing slightly different behaviors. In other words, based on the extended Gutzwiller approximation, two novel channels become emergent and the resulting four channels are renormalized by $g^{t}_{i\sigma}$ with lightly different weights;
\item[(iv)] We have checked that the terms in the last two rows in Equation (\ref{eq:gf}) correspond to a Cooper-pair breaker, and have no direct relations with the magnetic properties of the trilayer system;
\item[(v)] Since for our free energy function $\mathcal{F}$, we allow both ferro- and antiferro-type magnetic structures in $\Gamma^{\sigma}_{\alpha \beta}$, it shall be justified that there exist other identical channels of the proposed intramultilayer antiferromagnetic scattering when the AFM moments in one unit cell is ferromagnetically aligned. A quick calculation shows that the nearest-neighbor interlayer tunneling of single electron just provides such equivalent channels of intramultilayer AFM coupling if we start with $\mathop{\rm sgn}(\frac{m_{\alpha}}{m_{\beta}})\!=\!1$, therefore we can effectively use the strength $\kappa$ to denote all these interactions from the onsite to the nearest-neighbor single electron tunnelings between adjacent copper oxide layers and relax the magnetic structure to be either ferro- or antiferro-types, namely $\mathop{\rm sgn}(\frac{m_{\alpha}}{m_{\beta}})\!=\!\pm1$, in $\Gamma^{\sigma}_{\alpha \beta}$. This finalizes the form of the trilayer proximity model we studied;
\item[(vi)] Normally, the ferro-type magnetic structure is with suppressed AFM moments, whilst the coherent SC state carrying an inphase arrangement could greatly facilitate the $d$-wave pairing amongst all three layers in collaboration with an effective Cooper pair tunneling.
\end{enumerate}
We shall emphasize that all the above consequences are solely stemming from the strong electron correlation effects in underdoped high-$\textit{T}_\textit{c}$ cuprate superconductors.